# Filters for NIR astronomical photometry: comparison of commercial IRWG filters and designs using OpenFilters

ANWESH KUMAR MISHRA* and U. S. KAMATH

Indian Institute of Astrophysics, Bangalore 560034, India.
*Corresponding author. E-mail: mishraanwesh@gmail.com

**Abstract.** The photometric accuracy in the near-infrared (NIR) wavelength range (0.9–2.6 $\mu$m) is strongly affected by the variability of atmospheric transmission. The Infrared Working Group (IRWG) has recommended filters that help alleviate this issue and provide a common standard of NIR filtersets across different observatories. However, accurate implementation of these filters are yet to be available to astronomers. In the meantime, InGaAs based detectors have emerged as a viable option for small and medium telescopes. The present work explores the combination of IRWG filtersets with InGaAs detectors. A few commercially available filtersets that approximate the IRWG profile are compared. Design of more accurate IRWG filtersets suitable for the InGaAs sensitivity range is undertaken using an open-source filter design software – OpenFilters. Along with the photometric filters iZ, iJ and iH, design of a few useful narrow band filters is also presented. These filters present opportunities for small and medium telescopes for dedicated long-term observation of interesting infrared sources.

## 1. Introduction

Low-altitude astronomical sites house a significant number of small and medium aperture telescopes. Photometric observation of bright and variable stars – particularly in the near infrared (NIR: 0.9–2.6 $\mu$m) – is one of the areas where these telescopes can make a significant contribution. Presently most of NIR observation is being carried out by observatories located in high and dry sites. The primary science goal of these facilities are usually towards the fainter sources; and the instruments on these telescopes face saturation issues when observing bright infrared sources. The role of conducting dedicated long-term observing programs on such sources is best carried out by the small and medium sized telescopes. A large number of such telescopes, typical apertures ranging from 50 cm to 2 m, are present in low-altitude observatories. These facilities were extensively used in the era of photographic and photoelectric detectors when the observer played an active role in the telescope operation. With time, observational astronomy has moved towards large aperture telescopes at high-altitude sites and the demand for observing time on the smaller telescopes has reduced. Subsequently, a number of low-altitude observatories have been re-purposed as training facilities (e.g., Girawali Observatory[1]) or for outreach programs (e.g., Mt. John Observatory, NZ[2] and Purple Mountain Observatory). It has now become possible to obtain time on these facilities for dedicated and long-term observing programs on variable stars. Additionally, these sites have better ease of access compared to high-altitude sites and instruments can be developed for these telescopes within shorter time and lower cost. As these telescopes have limited weight carrying capability as well as modest operating budgets; there is a need for simple, low cost and easy to operate instrument that can achieve good photometric accuracy.

Recent technological progress holds promise for the feasibility of such an instrument. The first one is the availability of low-noise InGaAs detectors (Henden 2002; Sullivan *et al.* 2014). These detectors have good sensitivity in the 0.9–1.8 $\mu$m wavelength range and are not sensitive in the thermal infrared (>3.5 $\mu$m). Additionally, these detectors have minimal cooling requirements which can be achieved by thermoelectric cooling. The other significant progress is the introduction of the Infrared Working Group (IRWG) standard for NIR photometric filters (Milone & Young 2005). These filters promise to improve photometric accuracy from low-altitude sites and present opportunity for standardization of NIR filters across different observatories. The combination of InGaAs detectors with IRWG filterset allows for good photometric accuracy using simple and low-cost instruments. For InGaAs sensitive wavelength range, it is possible to use off-the-shelf glass lenses to design simple yet high throughput re-imaging optics (Mishra & Kamath 2021a). First light of such an instrument focusing on very bright infrared sources is discussed in Mishra & Kamath (2021b). In this work, we focus on the choice of IRWG equivalent filtersets from the point of view of small and medium telescopes. Commercially available filters that approximate the IRWG standard are compared for their suitability in replicating the recommended filter profile. Simple SNR calculations are done to estimate the necessary integration times to achieve high SNR while observing bright stars. Finally, multilayer filter design using OpenFilters (an opensource filter design tool) is presented for accurate implementation of the IRWG recommendation within the InGaAs sensitivity range.

---

[1] http://www.igo.iucaa.in/.
[2] https://www.darkskyproject.co.nz/.

## 2. Filters for NIR photometry

The atmospheric extinction in the near-infrared wavelengths is primarily defined by the absorption characteristics of molecules such as $H_2O$ and $CO_2$. The exact wavelength of observation and the line of sight concentrations of these molecules determine the amount of stellar flux that is absorbed (Bass 2010; Tokunaga *et al.* 2013). In such a scenario, the extinction is variable with respect to airmass (Manduca & Bell 1979) as well as molecular concentration. Of particular concern is water vapor which can vary in concentration even in short timescales. For this reason, high-altitude dry sites have traditionally been preferred for infrared astronomy and low-altitude sites which have inherently high water vapor concentrations are usually considered sub-optimal. The IRWG has recommended filtersets to alleviate issues related to extinction variability as well as to standardize infrared filtersets across different ground based observatories. Implementation of these filters is expected to result in better utilization of the small and medium telescopes that are present at low-altitude astronomical sites.

A simplified demonstration of the necessity of IRWG filtersets is presented in Figure 1. The atmospheric transmission of a low-altitude site (Kavalur – 750 m) and a high-altitude site (Hanle – 4500 m) are compared in Figure 1(a). The software ATRAN,[3] developed by Lord (1992) was used to estimate the atmospheric transmission. This software takes site parameters such as the latitude and altitude as input to estimate a model atmosphere. The line strength of major absorbing molecules are estimated by using the HITRAN[4] database. Finally, the atmospheric transmission is calculated by subtracting the total absorption contributions of all the molecules.

Specific values of precipitable water vapor (PWV) content can be given as input to ATRAN to estimate the variability of transmission. Maximum variability of transmission for a site can be approximated as:

$$V(\lambda) = T_{\text{high}}(\lambda) - T_{\text{low}}(\lambda), \quad (1)$$

where $T_{\text{high}}(\lambda)$ is transmission as a function of wavelength at 10% of maximum PWV of the site – indicative of a particularly dry night and high transmission values. $T_{\text{low}}(\lambda)$ is transmission as a function of wavelength at 90% maximum PWV – indicative of a particularly humid night with lower transmission.

This variability as a function of wavelength is shown in Figure 1(b). Two curves are drawn, one for Kavalur (ATRAN estimated PWV = 14.4 mm) and one for Hanle (ATRAN estimated PWV = 2.8 mm). PWV measurements have been carried out at the site of Hanle by Ananthasubramanian *et al.* (2004) and Ningombam *et al.* (2016). These measurements— excluding data from months which are affected by the monsoon – are in good agreement with the upper limit of 2.8 mm provided by ATRAN. In Figure 1(b), the variability of transmission is seen to be larger for the low-altitude site. However, it is also seen that the variability is much higher at the edges of the transmission windows compared to the centre. The infrared working group (IRWG) designed filters are optimized to avoid the regions that are affected the most due to variation of water vapor. The optimization methodology to reach at the exact filter pass-bands is discussed in Milone & Young (2005, 2008, 2011). The filter profiles of these improved filtersets are shown in

---

[3]https://atran.arc.nasa.gov/cgi-bin/atran/atran.cgi.
[4]https://hitran.org/.

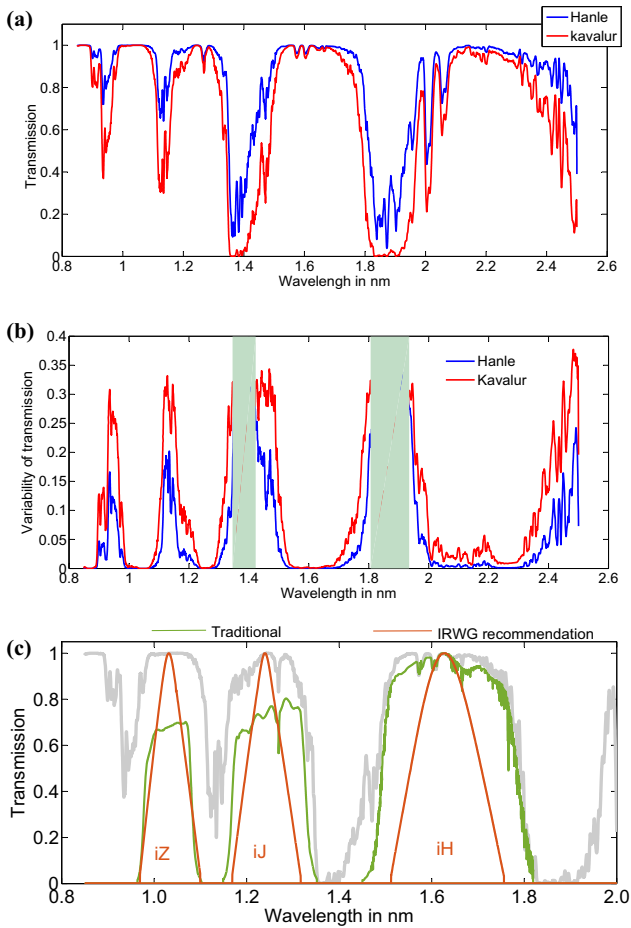

**Figure 1.** A demonstration on the necessity of IRWG recommended filtersets: transmission at a low-altitude site (Kavalur) and a high-altitude site (Hanle) was estimated using the software ATRAN (Lord 1992). The transmission at low-altitude sites is usually lower than high-altitude sites (a). At the same time the variability of transmission is also worse (b). The variation is particularly severe at the edges of the transmission windows. The red filter profiles in (c) are the IRWG recommended filtersets. The green curves are filter profiles from UKIRT. The J and H equivalent ones are derived from Johnson standards whereas the Y filter (IRWG equivalent iZ) is new as the original/extended Johnson standard did not have a filter in this wavelength range. Compared to Johnson derived filtersets (which usually have edges that are strongly affected by water vapor concentration), the IRWG filterset selectively rejects the wavelength regions that are strongly affected by water vapor.

Figure 1(c) in comparison to traditional filters. The IRWG filter sets are narrower in bandwidth and have their centre wavelength shifted slightly. The IRWG standard is also an opportunity to standardize filters across different observatories. Historically, after pioneering observations by Johnson (1965), Johnson *et al.* (1966), the development of NIR filtersets was undertaken independently by the major observatories. This has led to the existence of a large number of filter standards (Stephens & Leggett 2003), thereby making it difficult to compare infrared photometric results from different observatories. These filters were generally designed to allow for the maximum throughput by using a filter profile larger than the atmospheric transmission windows. In such cases, the exact filter transmission is determined by the atmospheric windows, which are variable in nature. As such, these filters are best used from high-altitude sites only. The IRWG recommended filter profiles are well within the atmospheric windows and are self-defined irrespective of the observatory altitude. These filters solve the variability of transmission issue as well as provide an opportunity for filter standardization across different observatories.

### 2.1 *Ideal and practical IRWG filtersets*

The IRWG standard is specified in detail by Milone & Young (2005) by specifying the transmission at multiple points. A triangular profile is considered to be ideal, however smooth curves joining these points with a relative flat tops are also acceptable as they are more easily realizable as multilayer thin film filters. The aim here is to avoid sharp points such as in a triangular/trapezoidal profiles. These smooth profiles for the IRWG filter bands iZ, iJ and iH are obtained by spline fitting as shown in Figure 2.

However, the unique shapes of these filters make it difficult to produce exact practical replica of these filters. Further, there is also the challenge of blocking of these filters in the out-of-band wavelength range. The IRWG definition requires for a broad blocking range over the complete sensitivity range of the detector, (upto 2.5 $\mu$m for HgCdTe, upto 5 $\mu$m for InSb and upto 1.8 $\mu$m for InGaAs). This is particularly important towards the longer wavelengths where the issue of thermal background can quickly overwhelm stellar sources (Milone & Young 2005). To our knowledge, practical filtersets from commercial suppliers are not yet available in the exact IRWG profile. There do exist filters that roughly match the centre wavelength and full width at half maximum (FWHM) of the IRWG filter set. These are available from companies such as OmegaFilters[5] and Custom Scientific[6]. Combining off-the-shelf short pass and long-pass filters from suppliers such as Edmund optics can also approximate the IRWG centre wavelength and FWHM. The exact short-pass and long-pass filters for this purpose are listed in Table 1. In cases where the filter definition by the manufacturer is not available for the complete wavelength range (InGaAs sensitive wavelengths range – discussed in detail in Section 3), a band extend filter is used to ensure the filter blocking is extended for the complete wavelength region of interest (Table 3). In Figure 2, transmission profiles of these practical filtersets are compared to that of the ideal IRWG profile.

---

[5]https://www.omegafilters.com/.

[6]https://customscientific.com/.

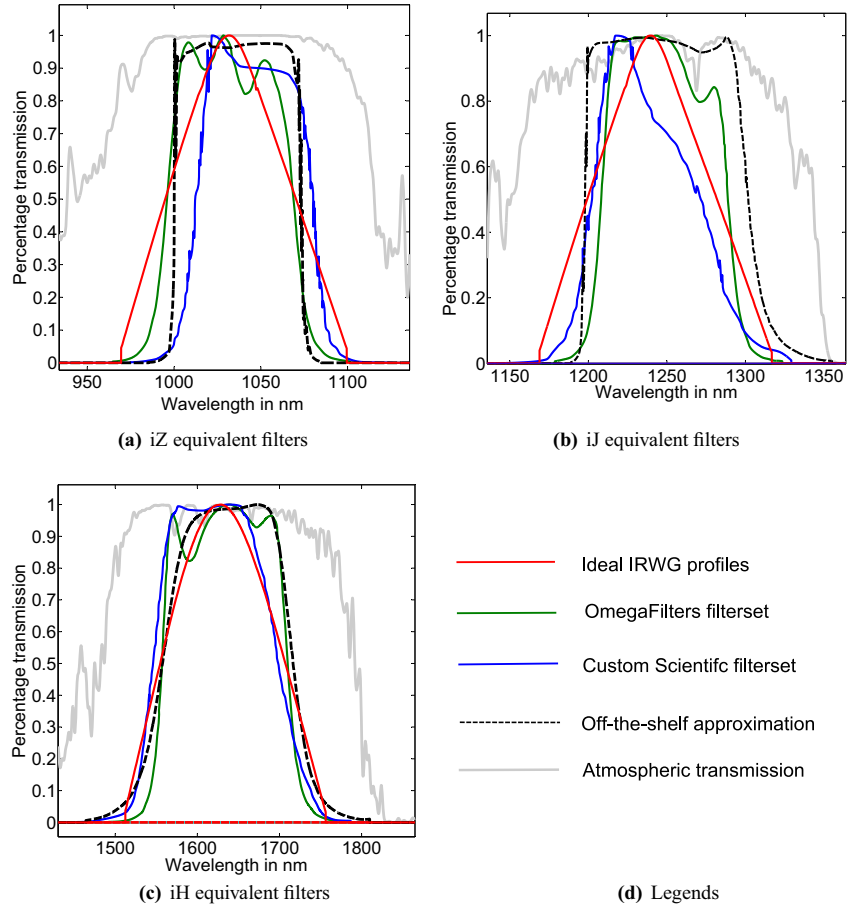

**Figure 2.** Examples of available filters for IRWG profiles: the ideal IRWG profile in red is compared to that of commercially available and off-the-shelf filters. The green filter profiles are an omega opticals filterset, the blue filter profiles are Custom scientific filters obtained from Milone & Young (2005). The imperfection in the profiles are due to the lower resolution of the source image. The Software Webplot Digitizer was used to obtain filter profiles from curves. The atmospheric transmission is shown in grey in the background. Although the commercial filters do not match the IRWG profile exactly, they do reproduce the centre wavelength and FWHM of these filters. It is of note that here we are prioritizing the shape of the filter as a selection criteria over the absolute transmission or throughput of the filters. This is done for two reasons: (1) interference filters can usually achieve peak transmission better than 90%, and hence it is difficult to compare their transmission curves with each other unless all filters are measured in the same physical setup and (2) the shape of the filter is the more important factor in providing immunity against variability of water vapor.

As none of the practical filters provide an exact match, there is need for a method to compare these filters in how well they replicate the IRWG profile. Comparing synthetic stellar magnitudes is a practical way to achieve this. For this purpose, 57 stars from the UKIRT bright list were selected. These stars are part of the MaunaKea primary standards and have their magnitudes known to an accuracy of 0.01 mag. A synthetic spectral energy distribution (SED) for each of these stars were obtained from SVO using the models of Coelho (2014). From this SED, and the known magnitudes in the UKIRT filter bands, the magnitude at any other filter profile can be calculated as:

$$M_{J|H} = -2.5 \times \log \frac{\int F_{\text{star}}(\lambda) \times S \times T_f(\lambda) \, d\lambda}{\int F_{\text{Vega}}(\lambda) \times T_f(\lambda) \, d\lambda}, \qquad (2)$$

**Table 1.** List of practical filters: A few practical implementation of the IRWG filtersets are compared to that of the ideal IRWG profile as well as traditional Johsnon derived filtersets. Filters with prefix 'i' (as in iJ) are ideal IRWG profiles. Filters that have prefix 'Ci' are from Custom Scientific. Filters with prefix 'Oi' are from OmegaFilters. Filter that have prefix 'Otsi' are off-the-shelf approximations using short-pass and long-pass filters.

| Filter | Description | $\lambda_c$ (nm) | FWHM (nm) |
|---|---|---|---|
| Y | – | 1032.5 | 103.4 |
| iZ | Ideal IRWG profile | 1032.8 | 73 |
| CiZ | Custom scientific profiles | ∼1042 | ∼70 |
| OiZ | OmegaFilters profile | ∼1030 | ∼70 |
| iZ* | Long-pass 1000 nm and short-pass 1075 nm (Edmund optics 84766 and 86118) | ∼1035 | ∼75 |
| J | – | 1252.8 | 159 |
| iJ | Ideal IRWG profile | 1240.0 | 79 |
| CiJ | Custom scientific Profile | ∼1230 | ∼70 |
| OiJ | OmegaFilters profile | ∼1245 | ∼80 |
| iJ* | Long-pass 1200 nm and short-pass 1700 nm (Edmund Optics 89666 and 84658) | ∼1250 | ∼100 |
| H | – | 1642.3 | 292 |
| iH | Ideal IRWG profile | 1628.0 | 152 |
| CiH | Custom scientific profile | ∼1620 | ∼150 |
| OiH | OmegaFilters profile | ∼1635 | ∼160 |
| iH* | Bandpass filter 1530–1730 nm (Spectrogon BBP-1530-1730) | ∼1630 | ∼180 |

where $M_{J|H}$ is the observed magnitude of the star, $F_{\text{Vega}}$ is the synthetic SED of the zeroth magnitude star, $F_{\text{star}}$ is the synthetic SED of the star Synthetic SEDS are based on models from Coelho (2014) obtained from the SVO[7]. Illustrative plot of the SEDs are shown in Figure 3(a). $T_f$ is the transmission profile of the filter (UKIRT J or H obtained from SVO[8], $S$ is a scalar multiplier by which the SED of the star needs to be scaled to produce the observed magnitude.

The above equation can be solved for $S$ and once $S$ is known, then the magnitude of the star in any other filter is calculated as:

$$M_{iZ|iJ|iH} = -2.5 \times \log \frac{\int F_{\text{star}}(\lambda) \times S \times T_{\text{if}}(\lambda)\,d\lambda}{\int F_{\text{Vega}}(\lambda) \times T_{\text{if}}(\lambda)\,d\lambda}, \tag{3}$$

where $M_{iZ|iJ|iH}$ is the magnitude in the new filter band, $T_{\text{if}}(\lambda)$ is the filter profile of the new filter bands (collected from Figure 2).

Using this method, the synthetic magnitudes were estimated for ideal as well as the practical IRWG profiles. A good approximation of the IRWG filterset will have to replicate both the profile and the centre wavelength of the IRWG filterset and hence will produce the same magnitude. Assuming that the magnitude estimate for the IRWG profile is the reference, error in each filter is simply the deviation in the estimated magnitude. The calculated error for about 57 standard stars are shown in Figure 3(b). The error in magnitude over a range of stellar SEDs serves as a parameter to evaluate the quality of these filters. The filterset constructed from off-the-shelf short-pass and long-pass filters is within 0.1 mag and both Omega Optical and Custom Scientific filtersets produce errors within 0.05 mag of the ideal filter profiles. Availability of IRWG equivalent filters is an important step in widespread implementation of these filter in astronomical instruments. A rudimentary discussion on filter transforms between various IRWG filtersets and between IRWG and Johnson filtersets is presented in Appendix B.

An alternative method to obtain these filters is by designing these profiles using a multilayer thin film stack. As discussed in previous sections, filters that roughly match the centre wavelength and the FWHM are now available commercially. Therefore, we shall focus our efforts on designing filters that match the IRWG recommendation more precisely. An exact profile of the IRWG filters will have the following benefits. First, the IRWG profile is designed to be more resilient against the variability of water vapor. The profiles are centred within the atmospheric window where the influence of water vapor is minimal and taper off with decreasing sensitivity at regions that have increasingly more influence of water vapor absorption. These profiles offer the optimum balance between achievable photometric accuracy and good throughput. Small deviations (within 5% of prescribed transmission values according to IRWG) can be tolerated, but larger deviations can compromise the ability of the filter to remain independent of the atmosphere, e.g., the spurious peak on the OmegaFilters 'OiJ' (green curve in Figure 2b) is off by more than 20%. At this point, it is also important to make a distinction between the shape profile and the peak transmission of the filter as defining characteristics for filter evaluation – even though, ideally, it is preferable to have both. For example, an exact profile with a peak transmission of 80% peak transmission may be preferable to a profile that has peak transmission of 85% but matches the profile poorly. This is because small throughput errors can be zeroed out in the standard photometric procedure whereby stellar magnitudes are referenced to standard stars with known brightness. However, variations due to water vapor are more sporadic in nature and can be difficult to remove. In this aspect, a more precise filter profile is desirable for achieving better photometric accuracy.

---
[7]http://svo2.cab.inta-csic.es/theory/newov2/index.php.
[8]http://svo2.cab.inta-csic.es/theory/fps/.

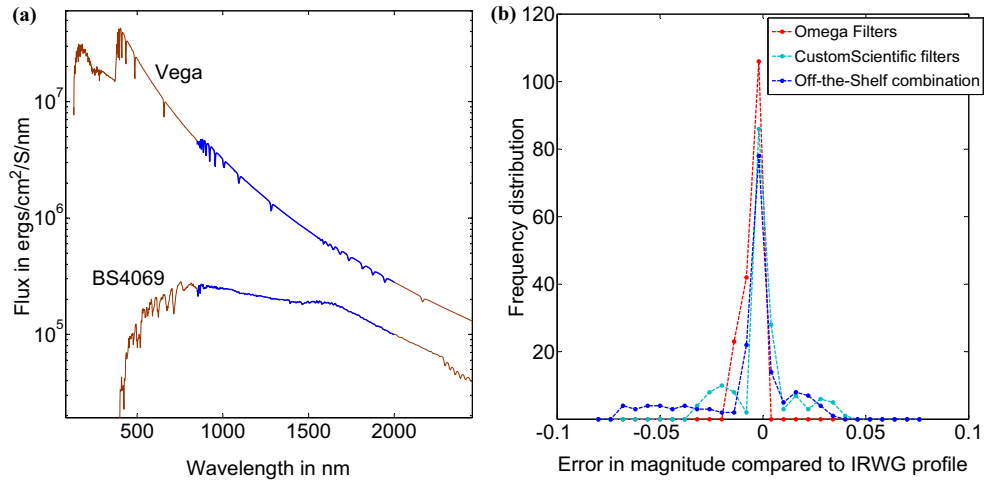

**Figure 3.** The filters were compared for magnitudes errors with 57 MaunaKea primary standards. SEDs collected from Coelho (2014) was used for this purpose. An example plot of SEDs of stars are shown in (a) with the SEDs within the detector sensitivity range marked in blue. The errors estimated for different filtersets are shown in (b). The comparison is between the filtersets and not the filters themselves; so the total number of sample points for each filterset in (b) are 171 (57 × 3). The practical filtersets match the IRWG profile within 0.05 mag and the off-the-shelf implementation matches the IRWG profile within 0.1 mag.

Second, having an exact filter is going to make observations from different telescopes and observatories easily comparable with each other. This will provide opportunity for a unified filter standard for different telescopes and observatories. Keeping these factors in mind, we shall explore the possibility of producing filter profiles that match the IRWG recommendation more closely than commercial filters. We shall use the open-source filter design software OpenFilters to explore the design complexity for photometric bands iZ, iJ and iH which fall within the InGaAs sensitivity range. The reason for limiting ourselves to InGaAs detectors is discussed in the following section.

## 3. InGaAs detectors for NIR photometry

Traditionally, InGaAs detectors have been used mostly for fibre-optic applications and are an emerging detector technology for low light level applications. The long wavelength response of most InGaAs detectors only extend upto 1.7 $\mu$m. With modification of the ratio of InAs and GaAs in the crystal structure, the long wavelength cutoff can be extended up to 2.5 $\mu$m – albeit with a corresponding increase in dark current. These detectors are not sensitive in the thermal infrared (>3 $\mu$m), have lower cooling requirements and can be operated with thermo-electric cooling only. Additionally, these detectors are also more accessible to the general astronomical community as there are less export restrictions on these.

**Table 2.** Required integration time in seconds to achieve a SNR = 100 by using various combination of detectors and filters. The included filters are the IRWG profile iZ, iJ, iH filters and one representative narrow band filter within each of the atmospheric windows. The single pixel detectors and the array based detectors are compared for different magnitudes as they aim at different science cases. The magnitudes listed are for traditional (Johnson derived) J and H filterbands. The Poisson limit for an ideal detector with no dark current is also included as a reference. The estimations are for the collecting area of a 1-m class telescope. The details of the detectors such as dark current and read noise are also included in the table.

| Filter | Hamamatsu G12181-203K single pixel (0.3 mm) NEP: $3.5 \times 10^{-15}$ W/$\sqrt{Hz}$ 7.5 mag | Teledyne-Judson J23TE4-3CN single pixel (0.25 mm) NEP: $4 \times 10^{-16}$ W/$\sqrt{Hz}$ 10.0 mag | Princeton Instr. NIRvana640 $640 \times 480$ (20 $\mu$) $N_{dark}$: 150 e$^-$/pixel/s $N_{read}$: 75 e$^-$/pixel 12.5 mag | PhotonEtc. ZephIR 1.7 s $640 \times 480$ (20 $\mu$) $N_{dark}$: 150 e$^-$/pixel/s $N_{read}$: 35 e$^-$/pixel 12.5 mag | Poisson limit 12.5 mag |
|---|---|---|---|---|---|
| iZ | 31 | 42 | 40 | 28 | 8 |
| He I | 734 | 950 | 510 | 440 | 40 |
| iJ | 29 | 36 | 28 | 18 | 6 |
| Pa $\beta$ | 745 | 970 | 330 | 300 | 33 |
| iH | 54 | 70 | 35 | 23 | 7 |
| Fe II | – | – | 800 | 820 | 55 |

Currently, astronomical grade InGaAs detectors are commercially available either in single pixel or as small arrays upto $640 \times 480$ pixels. The dark noise of these detectors are expected to further reduce as the technology matures for low light level applications (Vermeiren & Merken 2017). It has been shown by Sullivan *et al.* (2014) that it is possible to use present InGaAs detectors to achieve high SNR for brighter stars with short integration periods. Using the 0.6 m telescope of the Wallace observatory, they were able to achieve SNR higher than 100 observing a 9.4 mag star for an integration duration of 21 s. The performance of a few more InGaAs sensors are presented in Table 2. The detectors are evaluated by the required integration times to achieve SNR = 100. The filters for which the SNR is calculated are the broadband IRWG profiles (iZ, iJ, iH) and one representative narrowband filter within each of the atmospheric windows (He I, Pa $\beta$, Fe II). These filters are listed in the column 1 of Table 2. For SNR estimation, instead of SED models, we are using zeroth magnitude spectral irradiance values from (Zombeck 2006) in the corresponding wavelength range scaled to the desired magnitude. This is done so that the calculations are general in nature and independent of the star being observed. The calculations are for an aperture of 1 m telescope and a throughput of 20%. The detailed process of SNR estimating is listed in Appendix A. This SNR calculation is aimed to produce a rough estimation of observability.

The array detectors in Table 2 are compared for 12.5 mag. The single pixel detectors are compared for a different magnitude as these detectors cannot compete with array detectors in terms of sensitivity and noise but are nevertheless very useful in observing bright and variable stars with simple and low cost instruments. The table is an illustration of the range of various sources that can be observed using modern InGaAs detectors. The theoretical Poisson limit for observing same 12.5 mag star is also included for comparison.

In Table 2, we have used the required integration duration to achieve good SNR rather than bright or faint limits as our evaluation criteria. This approach has been taken because the exact limit of bright and faint sources that can be observed is difficult to estimate as both of these limits are affected strongly by practical factors. The brighter limit is affected by issues such as detector readout rate, processing speed of readout electronics, ADC resolution, scattering issues within the optical system, the availability and accuracy of neutral density filters, etc. The fainter limit is affected by telescope tracking accuracy, stability of dark and bias, variability of sky background, zenith angle of the source, etc. As we have focused on establishing the general importance of InGaAs detectors rather than on any particular telescope/instrument, we have listed out the time required for good SNR as our criteria. Given that it is possible to achieve high SNR (>100) within short integration times, an approximate rule-of-thumb maybe suggested for bright and faint limits as follows, for sources that are brighter than the target 12.5 mag by more than 5–6 mag (i.e., <6 mag) saturation is likely to be an issue, and similarly sources that are fainter than 12.5 mag by more than 5 mag (i.e., >17.5 mag) are likely to be too faint.

From Table 2, it seems feasible to achieve good SNR within about a minute of integration time for broadband and about 1000 s for the narrow-band filters. Facilities present at low-altitude sites will be best utilized in focused/long-term observations of brighter sources at high SNR rather than operation close to their sensitivity limits. InGaAs technology allows for simple, lightweight and easy-to-operate instruments for such cases. Therefore, in our discussion we shall attempt to recreate IRWG profiles iZ, iJ and iH that fall within the InGaAs sensitive range. Of course, the discussed filter design process itself and the software used are general in nature and not specific to InGaAs detectors, but we shall focus on filters tailored to have blocking range that match closely to InGaAs sensitive wavelengths, i.e., 0.95–1.85 $\mu$m.

## 4. Multilayer filter design using OpenFilters

The modern approach of designing filters is based on multi-layer interference filters. A discussion on necessary complexity for realizing IRWG profiles as well the design flow using an open-source filter design software is presented. The basic building block of such filters are alternate stacks of thin films which have alternately low and high refractive indices. The complexity of design and optimization of these filters is best addressed by using specialized software designed for this purpose. To design such filters, the OpenFilters software package was used. OpenFilters is an open-source tool for designing multi-layer thin film filters. This software was developed by the Functional Coating and Surface Engineering Laboratory (FCSEL) and is available as a free design resource from the website https://www.polymtl.ca/larfis/en/links. The software is open-source. This is beneficial for re-optimizing the designs for different coating/manufacturing technologies and allows for easier collaboration between groups designing these filters. These aspects will be beneficial towards the goal of a common standard of NIR filters.

A practical guide to install and use the software is provided by Larouche & Martinu (2008) and useful help in getting started is available in the website. The brief design process for a filter is described as follows:

- *Initial settings:* The initial settings include defining the substrate (typically, fused silica) front and back mediums (typically, void/air) and the wavelength resolution. A wavelength resolution of at least 1 nm is necessary for accurately defining photometric filters. The initial setting is to be done in the `Filter: Properties` menu.
- *Filter specification:* The process of designing a filter using OpenFilters starts by specifying the exact transmission/reflection curve. Data points of transmission need to be collected from various sources and a smooth filter profile needs to be constructed by spline interpolation. This is an important step as smooth profiles are practically easier to realize by means of multi-layer interference filters. This transmission profile can be given as input to the software from an external `.CSV` file. The file should contain two columns of data specifying the transmission value as a function of wavelength. A third column specifying the desired tolerance is optional. The stopband of the filter can be specified as an array input specifying start and stop wavelengths and the desired transmission throughout this wavelength range. The necessary commands for this operation are listed in 'Filter: Add target' menu.
- *Stack formula:* The filter optimization starts from a defined stack formula, such as

  $[HLH]^n$ or $[LHL]^n$,

  where $H$ and $L$ are representative layers of high and low refractive index layers, and $n$ is the stack repetition. We have used $TiO_2$ as the high refractive index material and $SiO_2$ as the low refractive index material. The stack repetition decides the initial number of layers before starting an optimization. This is an important parameter as too few number of layers will not be able to produce the desired filter profile and having too many layers can result in the optimization to be too slow or not converge at all. The necessary commands for this operation are available in `Filter:Stack Formula` menu. To reduce manufacturing complexity, it is desired to produce the optimal filter profile with as few layers as possible.

- *Optimization and tolerancing:* Once a desired filter transmission profile and stack formula is available, the optimization can be initiated. To start with, OpenFilters calculated the transmission curve of the initial stack as a function of wavelength. The Chi square error of the desired profile compared to the present profile is used as the merit function. The aim of the optimization is to minimize this error by generating the optimal thickness of layers that will result in the best transmission profile. The `Design:Optimize` menu contains the necessary commands for such operations. The optimization needs to be continued till a satisfied filter profile is reached. However, there is a possibility that the algorithm may get stuck inside a local minimum. In such cases, a needle impulse can be given to attempt a recovery.

After this, the design needs to be verified for a tolerance analysis – available as an option in `Preproduction`. This analysis determines whether the design is practical to implement. For this analysis, an allowable manufacturing tolerance of each layers is specified in terms of percentage (typically 0.5%) or layer thickness. The analysis produces a mean as well as worst case scenario of resulting profiles. The design can only be considered practical if these errors are within a certain limit.

The characteristics of the designed filter – such as transmission and reflection as a function of wavelength – can be plotted by using the `Analyse` menu. These characteristics as well as the designed front index profile can be exported by means of `Export` function.

### 4.1 *Practical design approaches*

The performance of a filter is evaluated by how well it matches the exact transmission in the pass-band and good blocking in the stopband. The stopband is simply defined as the wavelength region that is outside the passband but within the detector sensitivity range. For InGaAs arrays this range is typically 950–1850 nm. It was found to be challenging to do both using a single multilayer stack. In such a case, the possible practical design approaches are:

- *Using external off-the-shelf blocking filters:* The process of filter design can be simplified if the multilayer stack is only required to provide the transmission profile and the blocking is ensured by utilizing off-the-shelf short-pass and long-pass filters. An example of this process is shown in Figure 4. The designed multilayer filter produces the necessary filter profile within the passband and provides blocking only for a short wavelength range with fringing effects elsewhere within the detector sensitivity range. This fringing is suppressed by an additional bandpass blocking filter that has good transmission in the passband and very low transmission in the stopband. As these filters only produce the filter profile and need external short-pass/long-pass blocking filters, these filters can be designed with a smaller number of layers (about 26–28). The combination of a short-pass and a long-pass filter can be used to implement the blocking bandpass filter and thus fully define the filter over the complete wavelength range.

Such filters are available as catalog filters from suppliers such as Edmund optics, Thorlabs, Spectrogon, etc. Particular useful combinations of these filters that produce blocking range for the InGaAs sensitivity range are presented in separate filters need to be combined into one. These filters will require for a collimated beam section in the optical imaging chain and are therefore suitable for designs that make use of a collimator-camera type design.

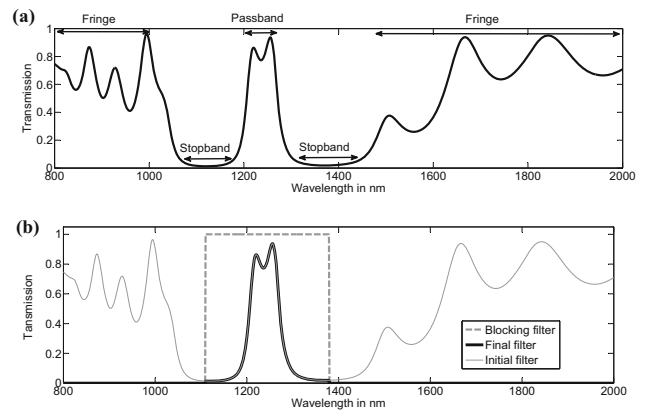

**Figure 4.** Two stage interference filter design: The filter design can be simplified into two stages. The exact filter transmission profile as well as a small blocking region is realized by means of a stack of interference filter consisting of alternating layers of high and low refractive index materials. The resulting transmission profile will be something like (a). Next, a bandpass filter is used to eliminate the regions of fringing over the complete sensitivity range of the detector – called stopband – to fully define the filter over the detector sensitivity range, as shown in (b).

**Table 3.** List of off-the-shelf filter combinations for blocking in InGaAs sensitive range: the filters starting with prefix EO are available from Edmund optics and those with prefix SP are available from Spectrogon.

| Filter | Long-pass | Short-pass | Band extend |
|---|---|---|---|
| iZ | EO-64708 (950 nm cut-on) | EO-89677 (1150 nm cut-off) | EO-84664 (1600 nm short-pass) |
| iJ | EO-84768 (1125 nm cut-on) | EO-84658 (1300 nm cut-off) | EO-84664 (1600 nm short-pass) |
| iH | EO-84686 (1475 nm cut-on) | SP-1845 (1800 nm cut-off) | EO-67301 (1300 nm long-pass) |

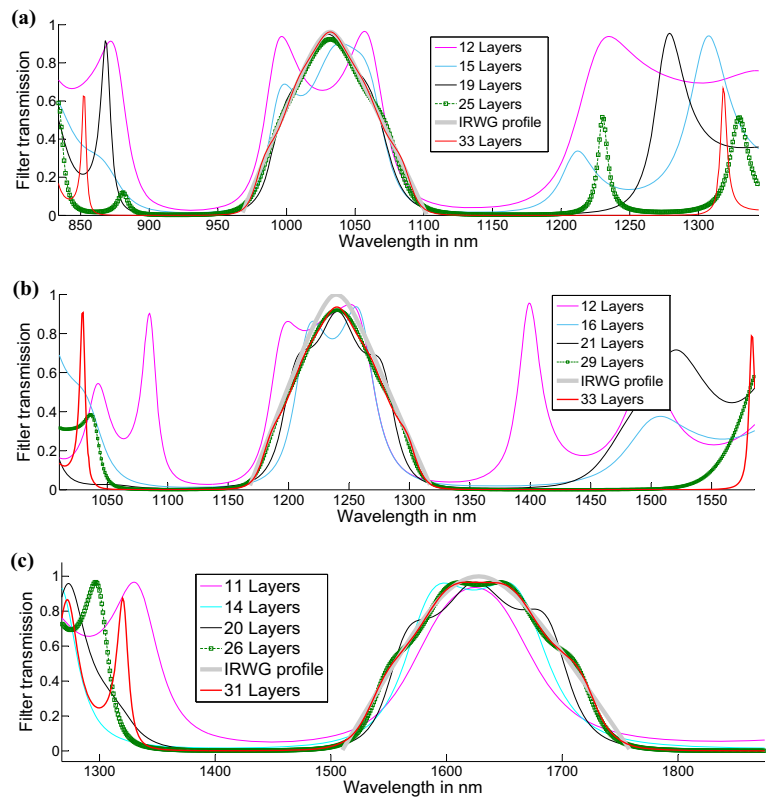

**Figure 5.** Required complexity of interference filters: the IRWG filter set profile is approximated by using interference filters. Across all filters – iZ, iJ and iH – higher number of layers equate to a better approximation (a), (b) and (c). About 26–28 layers are required to create a good approximation to the filter profile. These profiles when combined with the blocking filters shown in Table 3 will result in desired IRWG filters for the InGaAs sensitivity range. As these filters only replicate the filter profile without the blocking, these can be optimized with relatively smaller number of layers.

Table 3. These filters generally have good transmission (>95%), good blocking in stopband (minimum OD >2, typically >4) and are available in diameters ranging from one-inch to four-inch. The use of separate blocking filters allows the filter design to be focused on implementing just the transmission profile. Designed multilayer filters that produce IRWG profiles are presented in Figure 5. For each photometric filter, optimization was attempted for a range of multilayer stacks ranging from 12 to 30 layers. Good approximation for the filters as well as blocking in the vicinity of the passband was possible typically for 26 layers or more. When this transmission profile is combined with off-the-shelf blocking filters from Table 3, the resulting filters match the IRWG specification for the InGaAs sensitivity range. These filters, however, have the disadvantage of increased thickness as at least three

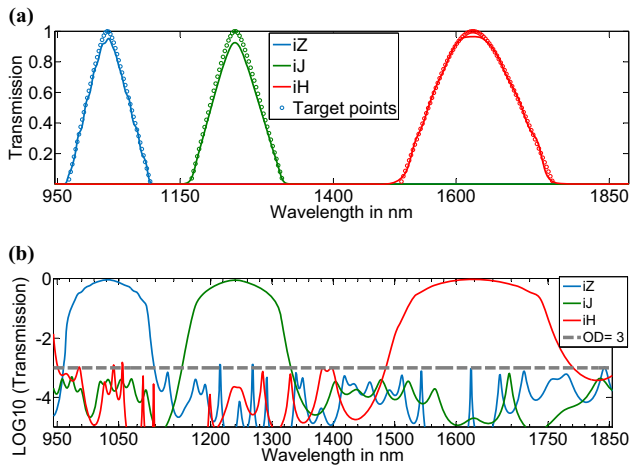

**Figure 6.** Profile and optical density of designed full stack filters: the transmission profiles of the designed filters is shown in comparison with IRWG target points in (a). The optical depth of the filters for the complete wavelength range is shown in (b). The designed filters achieve optical depth better than 3.0 for almost the complete wavelength range. To achieve both the filter profile as well blocking over the InGaAs sensitive wavelength range, about 70–80 layers were necessary.

- *'Full-stack' filter designs:* For applications where the extra thickness of these filters is not acceptable, a 'full-stack' design will be necessary. In this approach, the complete filter – both the transmission profile as well as the blocking range – is defined by a single multilayer stack. Design for a set of such filters are shown in Figure 6. The transmission profile is shown in Figure 6(a) and the blocking performance is shown in (b). The transmission profile matches the IRWG profile closely and the blocking range covering the InGaAs sensitivity range is achieved with an optical depth of 3.0. As these filters produce both the filter profile as well as the blocking range, these filters require higher number of layers to optimize. About 70–80 layers were necessary to produce the profiles shown. It is also of note that in this method it is difficult to tell how many layers within the filter contribute towards the filter profile and how many towards the blocking. The complete filter stack is optimized together to produce the desired transmission over the complete sensitivity range of the detector.

- *Design of narrow band filters:* Apart from the photometric filters, there also exist useful narrow band special purpose filters. White & Wing (1978) have presented methods for better spectral classification of late spectral type stars by using the absorption depths of VO and CN bands. The He I emission line is used as an indicator for the presence of chromosphere around cool stars (Spinrad & Wing 1969). Paschen $\beta$ and Fe II emission lines are useful diagnostic tools for shock induced variability in Mira type variables (Koo *et al.* 2016; Richter *et al.* 2003). $CH_4$ features are important while observing cool brown dwarfs and hot Jupiter like objects (Yurchenko *et al.* 2014). A number of narrow band filters and their continuum counterparts are listed in Table 4. These are special-purpose filters and will find use depending on specific cases. The narrow band filters can also be designed in a similar manner. The filter profiles were obtained from various sources as listed in Table 4. The transmission values as a function of wavelength was derived and set as targets within the OpenFilters. The optimization was carried out using a similar approach as before and the results in comparison to the target points are shown in Figure 7. The profiles designed here along with blocking filters as described in Table 3 can be used as complete implementation of these filters for InGaAs sensitive wavelength ranges.

One representative narrow band filter within each of the atmospheric windows was already included in Table 2 (the SNR of the other filters roughly scale with their FWHM). It is possible to get good SNR (>100) within about 1000 s of integration using these filters.

Steps were taken to ensure aid in the manufacturability of the filters. Layers with thickness smaller than 40 nm were excluded from the optimization. The Preproduction tool was used to simulate random errors in layer thickness. In Figure 8(a), the deviation for different percentage of random errors is shown for one example filter (iH filter from Figure 5b). A maximum error of 0.5–1.5% can be tolerated on the thickness of the layers if the exact profile is to be maintained without significant loss of transmission.

The effect of higher incidence angle on the filter profile is a shift of the filter profile towards shorter wavelengths. Figure 8(b) shows this shortward shift of the filter profile (iH filter from Figure 5b) as the

**Table 4.** List of narrow band filters: the specifications of a few narrow band filters were collected from literature. These data are used to generate target points for the OpenFilters software.

| Filter | $\lambda_c$ (nm) | FWHM (nm) | Reference |
| --- | --- | --- | --- |
| VO Cont. | 1039.5 | 5.0 | White & Wing (1978) |
| VO | 1054.0 | 6.0 | White & Wing (1978) |
| CN Cont. | 1081.0 | 6.0 | White & Wing (1978) |
| CN | 1097.5 | 7.0 | White & Wing (1978) |
| He I | 1085.0 | 15.0 | LBT LUCIFER He I |
| Paschen$\beta$ | 1287.2 | 14.5 | CFHT 5136 |
| J Cont. | 1218.8 | 15.5 | CFHT 6110 |
| O II | 1241.0 | 9.2 | CFHT 6113 |
| OH | 1189.2 | 11.4 | CFHT 8102 |
| Fe II | 1649.4 | 17.8 | CFHT 5202 |
| Fe II Cont. | 1700.7 | 14.2 | CFHT 5212 |
| $CO_2$ | 1625.7 | 70.5 | CFHT 5217 |
| $CH_4$On | 1691.9 | 105.2 | CFHT 8203 |
| $CH_4$Off | 1589.2 | 95.0 | CFHT 8204 |

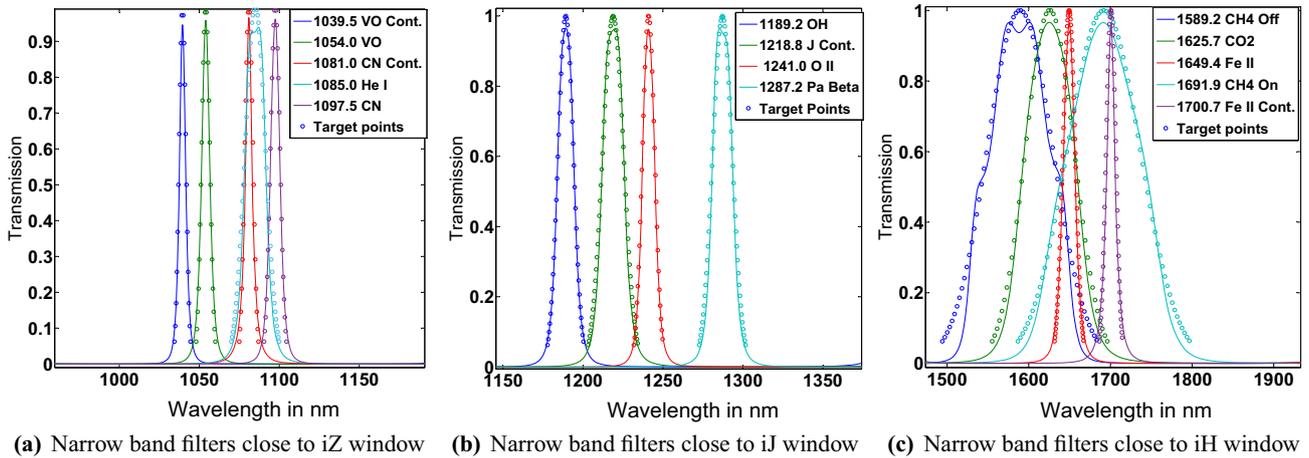

(a) Narrow band filters close to iZ window  (b) Narrow band filters close to iJ window  (c) Narrow band filters close to iH window

**Figure 7.** Realization of narrow band filters in OpenFilters: the target points for the narrow band filters were calculated from the specification in Table 4. The filters were designed using 26–30 layers in the OpenFilters software. The resulting filter profiles are compared with the target points as shown in (a), (b) and (c). These filter profiles will also need to be combined with the blocking filters from Table 3 to be used over the InGaAs wavelength range.

incidence angle is increased. We have found that upto 8° of incidence angle the filter profiles have minimal change. However, the present designs are mostly optimized for small field of view instruments keeping in mind the maximum size (640 × 512) of arrays. For larger arrays and larger field of view, a slight modification of the optimization process may be necessary, wherein the optimization target is moved slightly longward. These tolerance analysis are done to ensure that the final design would be implementable using the capabilities of modern ion beam or magnetron sputtering methods (Chen *et al.* 2020; Sakiew *et al.* 2020).

## 5. Conclusion

We have explored the combination of IRWG specification filters with InGaAs detectors from the point of view of small and medium telescopes. Practical implementations of IRWG equivalent filters were listed. The quality of these filters were evaluated by comparing magnitudes of 57 Mauna-Kea primary standards. The performance of these filtersets were found to be promising. We have also demonstrated that accurate filter profiles matching the IRWG specification can be designed using the open-source filter design software OpenFilters. The possibility of blocking using off-the-shelf shortpass and long-pass filter were explored as a practical option. A 'full-stack' design of the IRWG specified filters is also presented meeting a roughly optical depth of 3.0 for the InGaAs sensitive range. These filters are particularly interesting from the point of view of small and medium telescopes present at low-altitude astronomical sites. A SNR calculation using these filters along with various detector technologies was also presented to establish the usefulness of these filters for running dedicated observing programs on interesting infrared sources. We expect that with emerging science cases (such as variable stars, exoplanets, etc.), which require good photometric accuracy in NIR wavelength ranges, instruments that use IRWG filtersets will make significant contributions.

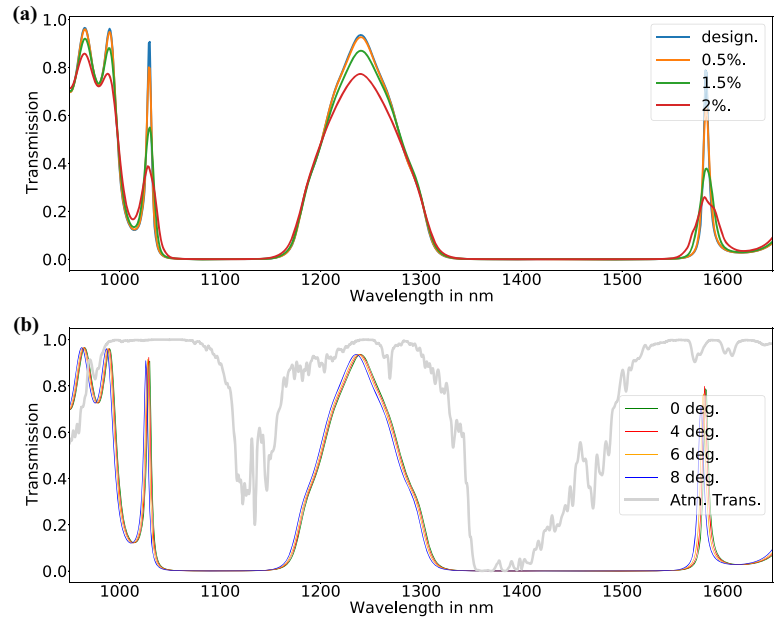

**Figure 8.** In (a) the degradation of the filter profile due to introduction of random errors in layer thickness is shown. Errors within 0.5–1.5% maybe tolerated without significant degradation of transmission. In (b) the shortward shift of the profile is shown with increase in incidence angle. For incidence angles upto 8°, the resulting change in transmission is minimal and within the allowable range. The atmospheric transmission window is shown in the background to demonstrate this. For large incidence angles (e.g., larger field of view) a slightly different approach to optimization may be taken (see text).

## Acknowledgments

The authors have made use of the WebPlotDigitizer tools to digitize data (https://automeris.io/WebPlotDigitizer). This research has made use of the SVO Filter Profile Service (http://svo2.cab.inta-csic.es/theory/fps/) supported from the Spanish MINECO through grant AYA2017-84089 (Rodrigo *et al.* 2012; Rodrigo & Solano 2020).

## Appendix A. Methods for SNR calculation

Collected flux in Watts from a star is specified as:

$$P_c = A_{\text{eff}} \times F_{\text{BW}} \times \Phi_{\text{zero}} \times 10^{(m_*/-2.5)}, \quad (A1)$$

where $A_{\text{eff}}$ is the effective collecting area of the telescope, $F_{\text{BW}}$ is the filter bandwidth in nm, $\Phi_{\text{zero}}$ is the flux from a zeroth magnitude star in W/m²/nm and $m_*$ is the magnitude of the star.

$F_{\text{BW}}$ and $m_*$ are known parameters and $\Phi_{\text{zero}}$ is collected from Zombeck (2006). The effective collecting area, $A_{\text{eff}}$ for a telescope is:

$$A_{\text{eff}} = \pi \left(\frac{D}{2}\right)^2 \times a \times b, \quad (A2)$$

where $D$ is the telescope diameter, $a$ is a factor corresponding to secondary obstruction; nominally 0.85 and $b$ is the throughput; nominally 0.2.

**Table 5.** Estimation of magnitudes of MaunaKea primary standards in IRWG equivalent filters. Filters with prefix 'i' (as in iJ) are ideal IRWG profiles. Filters that have prefix 'Ci' are from Custom Scientific. Filters with prefix 'Oi' are from OmegaFilters. Filter that have prefix 'Otsi' are off-the-shelf approximations using short-pass and long-pass filters.

| Star | iZ equivalents | | | | iJ equivalents | | | | iH equivalents | | | |
|---|---|---|---|---|---|---|---|---|---|---|---|---|
|  | iZ | oiZ | ciZ | otsiZ | iJ | oiJ | ciJ | otsiJ | iH | oiH | ciH | otsiH |
| BS337 | -0.38 | -0.39 | -0.41 | -0.39 | -0.90 | -0.92 | -0.90 | -0.97 | -1.74 | -1.75 | -1.70 | -1.71 |
| BS531 | 3.98 | 3.98 | 3.97 | 3.98 | 3.73 | 3.72 | 3.73 | 3.70 | 3.40 | 3.39 | 3.41 | 3.41 |
| BS696 | 5.41 | 5.41 | 5.41 | 5.41 | 5.47 | 5.47 | 5.47 | 5.47 | 5.53 | 5.53 | 5.53 | 5.53 |
| BS718 | 4.39 | 4.39 | 4.39 | 4.39 | 4.41 | 4.41 | 4.41 | 4.41 | 4.42 | 4.42 | 4.42 | 4.42 |
| BS1140 | 5.45 | 5.45 | 5.45 | 5.45 | 5.48 | 5.49 | 5.48 | 5.49 | 5.53 | 5.53 | 5.53 | 5.53 |
| BS1457 | -1.37 | -1.37 | -1.39 | -1.37 | -1.85 | -1.86 | -1.84 | -1.91 | -2.64 | -2.65 | -2.60 | -2.61 |
| BS1552 | 3.95 | 3.95 | 3.96 | 3.96 | 4.02 | 4.02 | 4.02 | 4.03 | 4.10 | 4.10 | 4.10 | 4.11 |
| BS1641 | 3.50 | 3.50 | 3.50 | 3.50 | 3.55 | 3.56 | 3.55 | 3.56 | 3.62 | 3.62 | 3.62 | 3.62 |
| BS1713 | 0.20 | 0.20 | 0.20 | 0.20 | 0.22 | 0.22 | 0.22 | 0.22 | 0.23 | 0.23 | 0.23 | 0.23 |
| BS1708 | -1.02 | -1.02 | -1.03 | -1.02 | -1.32 | -1.33 | -1.31 | -1.36 | -1.75 | -1.75 | -1.73 | -1.73 |
| BS2061 | -2.33 | -2.33 | -2.36 | -2.34 | -2.92 | -2.93 | -2.92 | -2.99 | -3.80 | -3.81 | -3.76 | -3.77 |
| BS2491 | -1.32 | -1.32 | -1.32 | -1.32 | -1.32 | -1.32 | -1.32 | -1.32 | -1.30 | -1.30 | -1.30 | -1.30 |
| BS2560 | 3.18 | 3.18 | 3.17 | 3.18 | 2.85 | 2.84 | 2.86 | 2.81 | 2.38 | 2.38 | 2.40 | 2.40 |
| BS2890 | 1.53 | 1.53 | 1.53 | 1.53 | 1.53 | 1.53 | 1.53 | 1.53 | 1.52 | 1.52 | 1.52 | 1.52 |
| BS2943 | -0.25 | -0.25 | -0.26 | -0.25 | -0.41 | -0.42 | -0.41 | -0.43 | -0.59 | -0.60 | -0.59 | -0.59 |
| BS2990 | -0.18 | -0.18 | -0.20 | -0.19 | -0.51 | -0.52 | -0.51 | -0.55 | -0.98 | -0.99 | -0.96 | -0.96 |
| BS3188 | 3.09 | 3.09 | 3.08 | 3.09 | 2.80 | 2.79 | 2.80 | 2.76 | 2.42 | 2.41 | 2.43 | 2.43 |
| BS3748 | 0.08 | 0.07 | 0.05 | 0.07 | -0.36 | -0.37 | -0.36 | -0.42 | -1.04 | -1.05 | -1.01 | -1.01 |
| BS3888 | 3.31 | 3.31 | 3.30 | 3.31 | 3.16 | 3.16 | 3.16 | 3.14 | 3.00 | 3.00 | 3.01 | 3.01 |
| BS3903 | 2.94 | 2.93 | 2.92 | 2.93 | 2.61 | 2.60 | 2.61 | 2.57 | 2.13 | 2.12 | 2.15 | 2.15 |
| BS3982 | 1.51 | 1.51 | 1.51 | 1.51 | 1.54 | 1.54 | 1.54 | 1.54 | 1.57 | 1.57 | 1.57 | 1.57 |
| BS4069 | 0.67 | 0.67 | 0.64 | 0.66 | 0.14 | 0.13 | 0.15 | 0.08 | -0.69 | -0.70 | -0.65 | -0.66 |
| BS4295 | 2.35 | 2.35 | 2.35 | 2.35 | 2.35 | 2.35 | 2.35 | 2.35 | 2.36 | 2.36 | 2.36 | 2.36 |
| BS4534 | 2.05 | 2.05 | 2.05 | 2.05 | 2.02 | 2.02 | 2.02 | 2.02 | 1.99 | 1.99 | 1.99 | 1.99 |
| BS4554 | 2.44 | 2.44 | 2.44 | 2.44 | 2.45 | 2.45 | 2.45 | 2.45 | 2.44 | 2.44 | 2.44 | 2.44 |
| BS4689 | 3.78 | 3.78 | 3.78 | 3.78 | 3.79 | 3.79 | 3.79 | 3.79 | 3.78 | 3.78 | 3.78 | 3.78 |
| BS4828 | 4.83 | 4.83 | 4.83 | 4.83 | 4.80 | 4.80 | 4.80 | 4.80 | 4.79 | 4.79 | 4.79 | 4.79 |
| BS4935 | 4.81 | 4.81 | 4.80 | 4.81 | 4.59 | 4.58 | 4.59 | 4.56 | 4.30 | 4.30 | 4.31 | 4.31 |
| BS4983 | 3.43 | 3.42 | 3.41 | 3.42 | 3.20 | 3.19 | 3.20 | 3.17 | 2.92 | 2.91 | 2.93 | 2.93 |
| BS5054 | 2.12 | 2.12 | 2.12 | 2.12 | 2.11 | 2.10 | 2.11 | 2.10 | 2.10 | 2.10 | 2.10 | 2.10 |
| BS5107 | 3.16 | 3.16 | 3.16 | 3.16 | 3.12 | 3.12 | 3.12 | 3.12 | 3.09 | 3.09 | 3.10 | 3.10 |
| BS5191 | 2.24 | 2.24 | 2.24 | 2.24 | 2.29 | 2.29 | 2.29 | 2.30 | 2.35 | 2.35 | 2.35 | 2.35 |
| BS5340 | -1.75 | -1.75 | -1.77 | -1.76 | -2.19 | -2.20 | -2.19 | -2.25 | -2.91 | -2.92 | -2.88 | -2.88 |
| BS5447 | 3.89 | 3.89 | 3.88 | 3.89 | 3.71 | 3.71 | 3.71 | 3.69 | 3.50 | 3.50 | 3.51 | 3.51 |
| BS5685 | 2.72 | 2.72 | 2.73 | 2.73 | 2.76 | 2.76 | 2.76 | 2.76 | 2.79 | 2.79 | 2.79 | 2.79 |
| BS5793 | 2.31 | 2.31 | 2.31 | 2.31 | 2.32 | 2.32 | 2.32 | 2.32 | 2.32 | 2.32 | 2.33 | 2.33 |
| BS6092 | 4.16 | 4.16 | 4.16 | 4.16 | 4.21 | 4.21 | 4.21 | 4.21 | 4.26 | 4.26 | 4.26 | 4.26 |
| BS6136 | 3.34 | 3.34 | 3.32 | 3.34 | 2.87 | 2.85 | 2.87 | 2.81 | 2.13 | 2.13 | 2.17 | 2.16 |
| BS6147 | 3.04 | 3.04 | 3.02 | 3.04 | 2.74 | 2.73 | 2.75 | 2.71 | 2.34 | 2.34 | 2.36 | 2.36 |
| BS6603 | 1.35 | 1.35 | 1.33 | 1.35 | 0.96 | 0.94 | 0.96 | 0.91 | 0.35 | 0.34 | 0.38 | 0.37 |
| BS6705 | 0.13 | 0.13 | 0.11 | 0.13 | -0.35 | -0.36 | -0.34 | -0.41 | -1.12 | -1.12 | -1.08 | -1.09 |
| BS6707 | 3.57 | 3.57 | 3.56 | 3.57 | 3.43 | 3.43 | 3.43 | 3.42 | 3.28 | 3.28 | 3.29 | 3.28 |
| BS7001 | 0.00 | 0.00 | 0.00 | 0.00 | 0.00 | 0.00 | 0.00 | 0.00 | 0.00 | 0.00 | 0.00 | 0.00 |
| BS7120 | 3.19 | 3.18 | 3.17 | 3.18 | 2.75 | 2.74 | 2.76 | 2.69 | 2.09 | 2.08 | 2.12 | 2.11 |
| BS7525 | 0.76 | 0.76 | 0.74 | 0.75 | 0.32 | 0.31 | 0.33 | 0.27 | -0.37 | -0.38 | -0.34 | -0.35 |
| BS7615 | 2.56 | 2.56 | 2.54 | 2.56 | 2.20 | 2.19 | 2.21 | 2.16 | 1.67 | 1.66 | 1.69 | 1.69 |
| BS7924 | 0.97 | 0.97 | 0.97 | 0.97 | 0.97 | 0.97 | 0.97 | 0.97 | 0.94 | 0.94 | 0.95 | 0.95 |
| BS7949 | 1.15 | 1.15 | 1.13 | 1.14 | 0.79 | 0.78 | 0.79 | 0.74 | 0.24 | 0.24 | 0.27 | 0.26 |
| BS8028 | 3.75 | 3.75 | 3.75 | 3.75 | 3.75 | 3.75 | 3.75 | 3.75 | 3.75 | 3.75 | 3.75 | 3.75 |
| BS8143 | 3.83 | 3.83 | 3.83 | 3.83 | 3.85 | 3.84 | 3.85 | 3.85 | 3.85 | 3.85 | 3.86 | 3.86 |
| BS8167 | 3.13 | 3.13 | 3.11 | 3.13 | 2.83 | 2.82 | 2.83 | 2.79 | 2.40 | 2.39 | 2.42 | 2.41 |
| BS8316 | 0.15 | 0.15 | 0.12 | 0.14 | -0.45 | -0.46 | -0.45 | -0.52 | -1.33 | -1.34 | -1.29 | -1.30 |
| BS8541 | 4.26 | 4.26 | 4.26 | 4.26 | 4.28 | 4.28 | 4.28 | 4.28 | 4.29 | 4.29 | 4.29 | 4.29 |
| BS8551 | 3.29 | 3.29 | 3.27 | 3.29 | 2.93 | 2.92 | 2.94 | 2.89 | 2.38 | 2.37 | 2.40 | 2.40 |
| BS8728 | 1.06 | 1.06 | 1.06 | 1.06 | 1.04 | 1.04 | 1.04 | 1.04 | 1.03 | 1.03 | 1.03 | 1.03 |
| BS8781 | 2.48 | 2.48 | 2.48 | 2.48 | 2.50 | 2.50 | 2.50 | 2.50 | 2.51 | 2.51 | 2.51 | 2.51 |
| BS8905 | 3.63 | 3.62 | 3.61 | 3.62 | 3.42 | 3.42 | 3.43 | 3.40 | 3.18 | 3.18 | 3.19 | 3.19 |

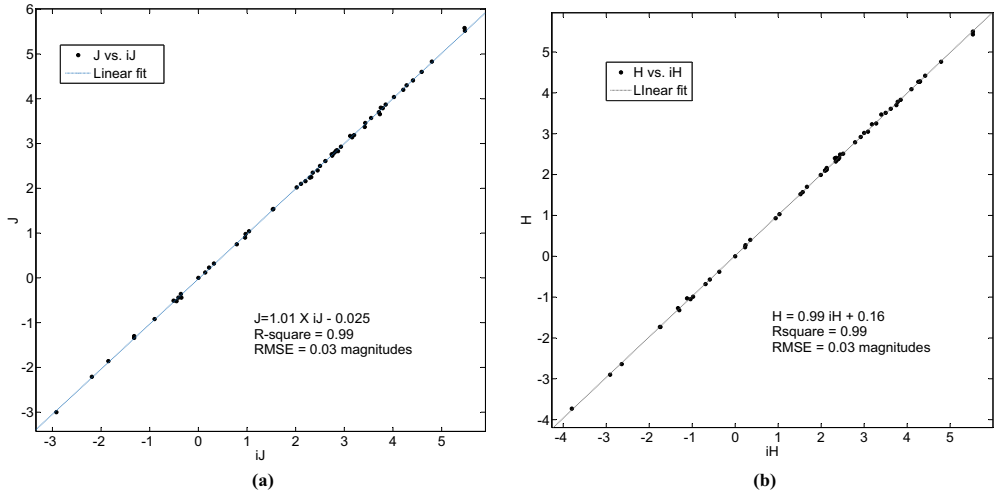

**Figure 9.** Using the synthetic photometry of the 57 MaunaKea primary standards, a rudimentary transform equations between IRWG filters iJ and iH can be constructed. A basic fit between J and iJ is shown in (a). A basic fit between H and iH is shown in (b). RMS error of 0.03 mag is seen in both the plots.

Using these parameters, the SNR for single pixel detectors can be calculated as:

$$\mathrm{SNR} = \frac{P_c}{\mathrm{NEP} \times \sqrt{1/T_i}}, \tag{A3}$$

where $P_c$ is incident energy in Watts, NEP is noise equivalent power of the detector in Watts/$\sqrt{\mathrm{Hz}}$ and $T_i$ is the on-source integration time.

SNR for array based detectors:

$$\mathrm{SNR} = \frac{P_c \times T_i}{\sqrt{P_c \times T_i + N_{\mathrm{dark}} \times p_n \times T_i + N_{\mathrm{read}}^2 \times p_n}}, \tag{A4}$$

where $N_{\mathrm{dark}}$ is the dark current of the detector given in e$^-$/pixel/S, $N_{\mathrm{read}}$ is the read noise of the detector given in e$^-$/pixel, $p_n$ is the number of pixels used to sample the stellar disk; nominally 16 and $T_i$ is the on-source integration time.

Using these relations between the desired SNR and time of integration, minimum integration time for SNR = 100 for various filter and detector combinations are given in Table 2.

## Appendix B. Filter transform between IRWG and Johnson filters

For the present typical achievable photometric accuracies in NIR (Milone & Young 2007; Wing et al. 2011), i.e., 3–5%, filter transforms between different implementations of the IRWG filterset may not be required. For more accurate photometry, i.e., 1% or lower, just 57 bright standards may not be sufficient for an exact transformation. We have included the synthetic photometric data of all 57 MaunaKea primary stars in various filtersets considered, should there be interest for such transforms. The data are included in Table 5 as a list of magnitudes. A rudimentary transform fit between IRWG and the Johnson filterset derived from the 57 MaunaKea standards is also included in Figure 9. For various aspects of filter transforms of IRWG filtersets, the work done by Milone & Young (2005) is to be referred